\documentclass[prb,aps,twocolumn,showpacs]{revtex4}
\usepackage{graphicx,color,psfrag,latexsym}
\usepackage{dcolumn}
\usepackage{amsmath,amssymb,epsf,bm}

\begin{document}
\title{Triplet supercurrent  in ferromagnetic Josephson junctions
by spin injection}
\author{A.~G. Mal'shukov$^{1}$
and Arne Brataas$^{2}$}
\affiliation{$^1$Institute of Spectroscopy, Russian Academy of
Sciences, 142190, Troitsk, Moscow oblast, Russia \\
$^2$Department of Physics, Norwegian University of Science and
Technology, NO-7491 Trondheim, Norway}
\begin{abstract}
We show that injecting nonequilibrium spins into the superconducting
leads strongly enhances the stationary Josephson current through a
superconductor-ferromagnet-superconductor junction. The resulting
long-range super-current through a ferromagnet is carried by triplet
Cooper pairs that are formed in s-wave superconductors by the
combined effects of spin injection and exchange interaction. We
quantify the exchange interaction in terms of Landau Fermi-liquid
factors.  The magnitude and direction of the long-range Josephson
current can be manipulated by varying the angles of  the injected
polarizations with respect to the magnetization in the ferromagnet.
\end{abstract}
\maketitle

\section{Introduction}

Studies of hybrid structures combining superconducting and
ferromagnetic components attract much attention due to their unique,
rich, and complex physical properties that are promising in a number
of potential applications \cite{Bergeret}.  The interface of an
s-wave superconductor with a ferromagnet is characterized by an
unusual proximity effect that is spatially oscillating and can lead
to a sign reversal of the critical current through
superconductor-ferromagnet-superconductor (SFS) Josephson junctions.
Such a reversal is equivalent to a $\pi$-shift in the current-phase
relation for the Josephson current. This interesting property is a
motivation for using the so-called $\pi$-junctions as elements of
superconducting quantum circuits for potential application in
quantum computing \cite{Feofanov}. However, the proximity effect in
ferromagnets does not reach far. Two critical tasks are to extend
its range and to find a way to manipulate the $\pi$-junction in
order to switch the device between its various phase states. In
contrast, Cooper-pairs can be transferred over relatively long
distances even in ferromagnets, if they are in a triplet state with
$\pm 1$ projections of their total spin onto the spin quantization
axis. Various mechanisms have been proposed that convert a singlet
pair into a triplet pair, such as a spatially dependent
magnetization \cite{Bergeret_triplet}, spin-flip scattering at FS
interfaces \cite{Eschrig}, and  precessing magnetization
\cite{Houzet}. A number of works in this direction has been
reviewed in Ref.\onlinecite{Bergeret}.

In this work, we will show that these tasks can be fulfilled via the
production and manipulation of a long-range proximity effect by
injecting spins into superconducting leads. The novelty  of our idea
is based on the important, and so far unaddressed, role played by
the electron-electron interaction in SFS. Our insight is that the
combined effects of spin-injection and electron-electron interaction
generate a long-range proximity effect despite the strong exchange
field in the ferromagnet. The conventional wisdom is that spin
polarized electrons can only exist as excitations in s-wave
superconductors, since the Cooper pairs do not carry a spin.
However, we will demonstrate that this simple picture, which is
based on the neglect of electron-electron interactions beyond
superconducting pairing correlations, misses qualitatively important
effects. Quantitatively, in simple metals, the exchange interaction
of itinerant carriers is noticeable and can be described in terms of
Landau Fermi-liquid factors. Although the exchange interaction does
not cause ferromagnetism in s-wave superconductors, it causes a
transfer of spin polarization from the quasi-particle excitations to
the condensate, in the form of polarized triplet Cooper pairs. When
such a triplet pairing is generated by the combined effects of
spin-injection and exchange interaction,  these pairs subsequently
tunnel through the ferromagnetic layer via  the long-range proximity
effect, if the spin polarizations in the leads and the layer are
not collinear. Only at this stage, which includes the so far unaddressed important electron-electron interaction, the situation becomes similar to
proposals of Ref.\onlinecite{Bergeret_triplet,Bergeret} where an
inhomogeneous magnetization gives rise to the long-range effect
provided by $\pm 1$ triplets. The relative angles between the spin
polarizations in the superconducting leads and in the ferromagnet
can be varied by controlling the injected spin
polarizations,  making it possible to vary the magnitude and sign of
the Josephson current. This enables manipulations of
$\pi$-junctions. In addition to the Josephson supercurrent, which is driven by the
difference in the condensate phases, there is also a  dissipative DC current.
The latter is induced by the spin polarization flow through the ferromagnetic layer with spin dependent conductivity. This dissipative current also can be manipulated by varying the injected polarization angles. As it will be shown, at some
angles it vanishes, so that the dissipative and supercurrents can be measured independently.

Various effects of an injected spin polarization and spin current on
the electric transport in SFS junctions \cite{MalshSFS,Bobkov} and
other superconducting systems
\cite{Maekawa,Takahashi,Morten,exper_injection} have been recently
considered. Despite this interest, the fact that the exchange
interaction transfers the spin polarization from the quasiparticles
to the condensate has not been addressed so far.

The article is organized by the following way. In Sec.II an
expression is derived connecting the triplet components of the
anomalous Green function to the nonequilibrium spin polarization in
superconducting leads. In Sec.III the Josephson and dissipative
currents are calculated. Finally, our results are discussed in Sec.IV.

\begin{figure}[tp]
\includegraphics[width=6cm]{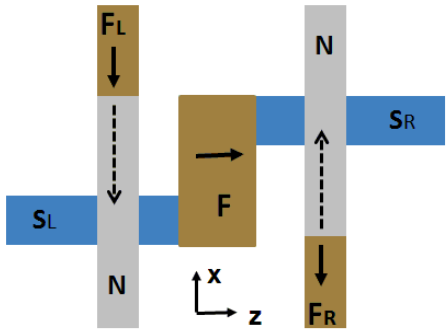}
\caption{(Color online) A sketch of the system. The electric current
flows in normal leads N through contacts with ferromagnetic leads
F$_L$ and F$_R$. Spin density is  injected from F$_L$ and F$_R$ into
N and further penetrates across tunneling barriers into
superconductors S$_L$ and S$_R$. The Josephson current flows between
these leads through a ferromagnetic layer F. Arrows show possible
magnetizations of the ferromagnets.} \label{fig1}
\end{figure}

\section{Triplet electron pairing function induced by spin injection}

How to efficiently inject a spin polarization into paramagnetic
metals is well known \cite{Jedema}. A  nonequilibrium spin
accumulation is induced by the electric current through a
paramagnetic-ferromagnetic interface. We consider the scenario that
the spin polarization further diffuses from a paramagnet through a
resistive barrier into a superconducting lead, so that the electric
circuit where the spin injection takes place is effectively
separated from the superconducting circuit. We assume that the
steady state spin polarizations are  generated in both
superconducting leads, in the vicinity of the F-layer. The sketch of
the system is shown in Fig.1. For clarity, we simplify the problem
by assuming that the FS contacts contain a barrier, so that the
proximity effect is weak. We also assume that  the spin relaxation
time $\tau_{\text{spin}}$ in the leads is long, so that the spin
diffusion length $l_\text{spin}$ is large compared to the SN contact
sizes and the coherence length. Consequently,  the spin densities
$\mathbf{s}_{L(R)}$ and the order parameters $\Delta_{L(R)}$ only
vary slowly in space near the left ($L$) and right ($R$) contacts.

The electronic transport through an SFS system, whose characteristic
dimensions are larger than the elastic mean free path, can be
described in terms of Usadel equations for angular averaged Green's
functions $g$ (for a review see \cite{Rammer}). These functions are
matrices in the Keldysh, spin, and Nambu spaces. We choose the spin
and Nambu spaces so that the one-particle destruction operators are
$c_{1\mathbf{k}\uparrow}=c_{\mathbf{k}\uparrow}$,
$c_{1\mathbf{k}\downarrow}=c_{\mathbf{k}\downarrow}$,
$c_{2\mathbf{k}\uparrow}=c^{\dag}_{-\mathbf{k}\downarrow}$,
$c_{2\mathbf{k}\downarrow}=-c^{\dag}_{-\mathbf{k}\uparrow}$, where
the labels 1 and 2 denote the Nambu spinor components, while
$\uparrow$ and $\downarrow$  are the spin indices.  The Keldysh
component $g^K$ of the Green function can be represented as
\cite{Rammer}
\begin{equation}\label{GK}
g^K=g^r h-hg^a \,,
\end{equation}
where $g^r$ and $g^a$ are the retarded and advanced functions,
respectively, and the distribution function $h$ is a diagonal matrix
in the Nambu space.

In order to determine the distribution $h$ in the superconducting
leads, the interfaces between these leads and the spin-polarized
normal metals must be considered. We use standard boundary
conditions relating fluxes through S-N (S-F) interfaces to Green
functions in superconductors and normal metals (ferromagnets). It is
assumed that the spin relaxation rates in the superconducting leads
are slow enough ($l_{\text{spin}} \gg r_{sn}\sigma_s$ ) and the
leakage of the spin polarization through the SF boundary is
sufficiently slow $r_{sn}/A_{sn} \ll r_{sf}/A_{sf}$, where
$1/r_{sn}$  and $1/r_{sf}$ are the interface conductances (per unit
square) of SN and SF interfaces, $A_{sn}$ and  $A_{sf}$ are the SN
and SF contact areas, and $\sigma_s$ is the normal-state
conductivity of the superconductor's lead. With these assumptions,
the distribution functions in the superconductor, $h^{(s)}$, and
normal metal, $h^{(n)}$, are equal to each other, $h^{(s)}=h^{(n)}$.
We further assume that nonequilibrium spins in N-leads are
thermalized with chemical potentials $\mu_{\uparrow}$ and
$\mu_{\downarrow}$ for the two spin directions. Therefore, denoting
by the subscripts 11 and 22 the corresponding matrix elements in the
Nambu space, we get for $h_{\uparrow(\downarrow)}\equiv
h^{(s)}_{11\uparrow(\downarrow)} = h^{(n)}_{11\uparrow(\downarrow)}$
and $\bar{h}_{\uparrow(\downarrow)}\equiv
h^{(s)}_{22\uparrow(\downarrow)} = h^{(n)}_{22\uparrow(\downarrow)}$
\begin{equation}\label{h}
h_{\uparrow(\downarrow)}=\bar{h}_{\uparrow(\downarrow)}=
\tanh\frac{\omega-\mu_{\uparrow(\downarrow)}}{2k_BT} \, .
\end{equation}
At the same time, the retarded ($g^r$) and advanced ($g^a$) Green
functions have the same forms as in an equilibrium superconductor.

Our calculation so far re-iterates the conventional wisdom of
spin-injection in superconductors: the effects are limited to a
spin-dependent statistical distribution function, while the retarded
and advanced Green functions do not change. In this picture, spin
injection does not lead to  the appearance of triplet correlations
in the condensate wave-function, which would cause long-range
Josephson tunneling through a ferromagnetic layer. Fortunately,
there is a mechanism to generate triplet correlations in
spin-polarized superconducting leads, which others have so far
overlooked. The electron-electron exchange interaction provides a
coupling between a spin accumulation and the spectral properties of
superconductors, in that spin polarized quasiparticles produce an
effective Zeeman field. The latter, in its turn, gives rise to
\emph{triplet} correlations that are described via the corresponding
spin components of the anomalous functions $g^r_{12}$ and
$g^a_{12}$. In Fermi-liquid theory, the effective Zeeman energy is
$\epsilon_{xc}(\bm{\sigma}\mathbf{N})$, where $\mathbf{N}$ is a unit
vector parallel to the injected spin polarization
$\mathbf{S}=\mathbf{N}S$ and
\begin{equation}\label{exc}
\epsilon_{xc}=\mathrm{G} S/2N_F \, .
\end{equation}
The spin-accumulation magnitude is
\begin{equation}\label{S}
S=-\frac{N_F}{4(1+\mathrm{G})}\int d\omega \text{Tr}
[\frac{(1+\tau_3)}{2}\sigma_z g^K]\,,
\end{equation}
where  $\tau_3$ and $\sigma_z$ are the Pauli matrices acting in the
Nambu and spin spaces, respectively, and $N_F$ is the density of
states at the Fermi level.  The renormalization factor
$1/(1+\mathrm{G})$, where $\mathrm{G}$ is the exchange Landau-Fermi
liquid parameter, appears when the spin-density of Eq. \ref{S} is
expressed in terms of a semiclassical Green function integrated over
energy \cite{Serene}. This factor is not qualitatively important in
our case, since $\mathrm{G}$ is not too close to the paramagnet
instability $\mathrm{G}=-1$. \cite{Landau} The exchange Coulomb
interaction in metals gives rise to a negative $\mathrm{G}$. For
example, the calculated value is -0.17 in Al \cite{Leiro}.
The spin density (\ref{S}) strongly depends on temperature,
mostly via the temperature dependence of the superconducting gap in
the energy spectrum. In order to determine $S$ and $\Delta$ in both
leads, Eq. (\ref{S}) have to be solved together with the
$S$-depended selfconsistency equation for $\Delta$.\cite{Takahashi}

Via the effective Zeeman energy of Eq.\ref{exc},
the retarded and advanced Green functions become spin-dependent.
\cite{Bergeret} Indeed, choosing the quantization axis along
$\mathbf{N}$, the anomalous functions
$f^r_{\uparrow\downarrow}=g^r_{12\uparrow\uparrow}$ and
$f^r_{\downarrow\uparrow}=-g^r_{12\downarrow\downarrow}$ become
\begin{equation}\label{F}
f^r_{\uparrow\downarrow(\downarrow\uparrow)}=\pm
\frac{|\Delta|\exp(i\phi)}{\sqrt{(\omega\mp\epsilon_{xc}+i\delta)^2-|\Delta|^2}}\,,
\end{equation}
where the phase $\phi$ of the order parameter $\Delta$ equals
$\phi_{L}$ and $\phi_{R}$ at the left and right contacts,
respectively.  The triplet component of this function with
0-spin-projection onto the $z$-axis is
$f^r_0=(f^r_{\uparrow\downarrow}+f^r_{\downarrow\uparrow})/\sqrt{2}$,
while the triplet components with $\pm$1-projections vanish,
$f^r_{\pm 1}=f^r_{\uparrow\uparrow(\downarrow\downarrow)}=0$. The
advanced function, as well as the conjugated functions $f^{\dag}$,
are determined from symmetry relations.

It is more transparent to discuss the Green functions in a basis
where the spin quantization axis is parallel to the magnetization in
the ferromagnetic layer, which is along $z$, as shown in Fig. 1. We
assume that  the spin polarizations in the left and right leads  are
rotated with respect to this axis by the angles $\theta_L$ and
$\theta_R$, respectively. We follow the convention that the three
components of the triplet $f_0,f_1,f_{-1}$ are related to a 3D
vector $\mathbf{a}=(a_x,a_y,a_z)$ with $a_z=f_0$,
$a_x=(f_{-1}-f_1)/\sqrt{2}$ and  $a_y=i(f_{-1}+f_1)/\sqrt{2}$
\cite{Landau2}. Hence, in the geometry shown in Fig. 1, after a
rotation of  $\mathbf{a}$ around the $y$-axis, we get in the new
basis $f^{\prime}_0=f_0 \cos\theta$ and
$f^{\prime}_1=-f^{\prime}_{-1}=-f_0 \sin\theta/\sqrt{2}$. So, by
using Eq. (\ref{F}) the triplet components in the left and right
superconducting leads are
\begin{equation}\label{FRL}
f_{\pm 1 R(L)}=-\frac{\sin\theta_{R(L)}}{2}(f_{\uparrow\downarrow}+
f_{\downarrow\uparrow})\,,
\end{equation}
where the labels $r$ and $a$ have been omitted from here and the
same magnitudes of $\epsilon_{xc}$ are assumed in both leads. In the
new basis, the distribution function (\ref{h}) is
\begin{eqnarray}\label{h2}
h_{L(R)}&=&h_{\uparrow}\frac{(1+\sigma_z\cos\theta_{L(R)})}{2}+
h_{\downarrow}\frac{(1-\sigma_z\cos\theta_{L(R)})}{2}+\nonumber \\
&&\sigma_y\sin\theta_{L(R)})\frac{h_{\uparrow}-h_{\downarrow}}{2}\,.
\end{eqnarray}

\section{The Josephson and dissipative currents}

What we have established is that the superconducting leads acquire
triplet pairing correlations determined by non-equilibrium spin
polarizations whose directions are tilted with respect to the
ferromagnet's magnetization in the SFS junction. We will show that
the current through such a triplet pairing-ferromagnet-triplet
pairing system consists of two parts, a dissipative contribution
controlled by the non-equilibrium distribution of spins in the
device, and a super-current driven by the phase difference between
superconductors and provided by the triplet components of the
superconducting condensates in the left and right leads.

Let us first consider the dissipative current. It can be expressed
in terms of the distribution function $h_\text{f}$ inside the
ferromagnet. Due to precession in the exchange field $B_\text{ex}$,
the spins that are not parallel to it decay quickly on the
length-scale $\sqrt{D_\text{f}/B_\text{ex}}$, where $D_\text{f}$ is
the diffusion constant. Therefore, only the components of
$h_\text{f}$ that are parallel and anti-parallel to $z$, denoted as
$h_{\text{f}\uparrow}$ and $h_{\text{f}\downarrow}$, remain finite
inside the ferromagnet, if the junction length $L \gg
\sqrt{D_\text{f}/B_\text{ex}}$. When the spin relaxation length is
larger than $L$, in the linear approximation these collinear
components satisfy the spin-conserving diffusion equation
$D_{\text{f}\sigma} \nabla^2_x h_{\text{f}\sigma}=0$, where
$\sigma=\uparrow,\downarrow$, that takes into account spin-dependent
diffusion coefficient in a strong ferromagnet. The solution of this
equation is a linear function of $x$ whose slope is obtained from
the boundary conditions  $\mp
r_{s\text{f}\sigma}\sigma_{\text{f}\sigma} \nabla_x
h_{\text{f}\sigma}|_{x=x_{L(R)}}=h_{L(R)\sigma}-h_{\text{f}\sigma}|_{x=x_{L(R)}}$,
where $h_{L(R)\sigma}$ are given by the first two terms of Eq.
(\ref{h2}). Taking into account that $r_{s\text{f}\sigma}$ and
$\sigma_{\text{f}\sigma}$ can depend on the electron spin and
assuming equal barrier transmittances at L and R contacts we obtain
\begin{equation}\label{hf}
h_{\text{f}\sigma}=\frac{h_{R\sigma}+h_{L\sigma}}{2}+\frac{h_{R\sigma}-h_{L\sigma}}{1+2\gamma_{\sigma}}\frac{x}{L}\,,
\end{equation}
where $x_{R(L)}=\pm L/2$ and
$\gamma_{\sigma}=(r_{s\text{f}\sigma}\sigma_{f\sigma}/L) \gg 1$.
Using Eqs. (\ref{hf}) and (\ref{h2}) we compute the dissipative part
of the current through the junction:
\begin{eqnarray}\label{jd}
&&j_d=\sum_{\sigma} \int d\omega \sigma_{\text{f}\sigma}\nabla_x
h_{\text{f}\sigma}=\nonumber \\
&&\frac{\delta\mu}{eL}\left(\frac{\sigma_{\text{f}\uparrow}}{1+2\gamma_{\uparrow}}-\frac{\sigma_{\text{f}\downarrow}}{1+2\gamma_{\downarrow}}\right)
(\cos\theta_{R}-\cos\theta_{L})
\,.
\end{eqnarray}
This current is proportional to the difference in the spin-up and
spin-down conductances
$(2r_{s\text{f}\sigma}+L/\sigma_{\text{f}\sigma})^{-1}$ of the total
ferromagnetic layer, including the interfaces; this is the well
known \cite{Aronov} connection between spin and electric transport
in ferromagnets. The electric current attains its maximum when
$\cos\theta_R=-\cos\theta_L=\pm 1$, and vanishes at
$\theta_R=\theta_L$, as well as at $\theta_R, \theta_L=\pm \pi/2$.
Such an angular dependence has a simple physical explanation.
The electric current (\ref{jd}) is proportional to the spin-current
through the junction. The latter attains its maximum when the nonequilibrium spin polarizations in the ferromagnetic layers are oppositely directed and it vanishes if these polarizations are collinear and have equal magnitudes. The spin
current obviously also vanishes if these polarizations are
perpendicular to the ferromagnetic magnetization axis, since
perpendicular components do not penetrate deep into ferromagnet. The
spin flow through the junction is accompanied by energy
dissipation. It is determined by the Ohmic losses in the ferromagnet
during transport of spin polarized electrons between the leads
having spin dependent electrochemical potentials. The dissipative
current of Eq. (\ref{jd}) is independent of the superconducting
phases $\phi_L$ and $\phi_R$. We assume that the electric potentials
of both contacts are equal. If the load is present in the circuit,
the spin current will induce a voltage difference. The latter, in
its turn, can cause periodic oscillations of the Josephson current
\cite{MalshSFS}.

When  $\sqrt{D_\text{f}/B_\text{ex}}$ is much shorter than the
junction length $L$ and the coherence length, the up and down-spin
Fermi surfaces become decoupled. In this regime, the supercurrent
$j_s$ through the junction  is determined by the decoupled tunneling
of $\pm 1$ triplet Cooper pairs at their respective ferromagnet's
Fermi surfaces. Unlike the dissipative current, the spin-dependence
of  the electron diffusion coefficients and conductivities is not so
important, at least in the case when the $B_{\text{ex}}\ll E_F$.
Therefore, in the leading approximation we set
$D_{\text{f}\uparrow}=D_{\text{f}\downarrow}=D_\text{f}$, and a
similar relation for the conductivities. Furthermore, in the linear
approximation, only the first term of  Eq. (\ref{hf}) has to be
taken into account. Moreover, since  the Josephson current is
determined by part of the distribution function that is odd in
frequency, from Eqs. (\ref{hf}), (\ref{h2}) and (\ref{h}) only the
spin-independent part $h_{\text{f}\uparrow}+h_{\text{f}\downarrow}$
contributes to the current. It is given  by
\begin{eqnarray}\label{js}
j_s&=&\frac{\sigma_\text{f}}{16e}\sum_{m=\pm 1}\int
d\omega\left[(f^r_{m} \nabla_x f^{r\dag}_{m} -  \nabla_x f^r_{m}
f^{r\dag}_{m})- \right. \nonumber \\ &&\left.(f^r \rightarrow
f^a)\right]\frac{(h_{\text{f}\uparrow}+h_{\text{f}\downarrow})}{2}\,,
\end{eqnarray}
where $f^{r}_{m}$ ($m=\pm 1$) are the retarded triplet components of
the anomalous function in ferromagnet and
$f^{a}_{m}(\omega)=f^{r}_{m}(-\omega)$, while
$f^{r\dag}_{m}(\omega)=f^{r*}_{m}(-\omega)$. When the
spin-relaxation length is much larger than $L$, and within the
linearized approximation, $f^{r}_{\pm 1}$ obey \cite{Bergeret}
\begin{equation}\label{F1}
D_{\text{f}}\nabla_x^2 f_{m} + 2i\omega f_{m}=0\,,
\end{equation}
with the boundary conditions \cite{Kupriyanov}
$r_{s\text{f}}\sigma_{\text{f}} \nabla_x f_{m}|_{x=x_{R(L)}}=\pm
f_{m R(L)}$, where $f_{m R(L)}$ are given by Eqs.
(\ref{F}-\ref{FRL}). After transforming the integral in Eq.
(\ref{js}) into a sum over the frequencies $\omega_n=\pi k_B T
(2n+1)$, $j_s$ can be finally represented in the form
\begin{equation}\label{js2}
j_s=\sin(\phi_L-\phi_R) \frac{L\sin\theta_R\sin\theta_L}{e\sigma_{\text{f}} R^2_{s\text{f}}} K \,,
\end{equation}
where
\begin{eqnarray}\label{K}
K=|\Delta|^2 k_BT\sum_{\omega_n > 0,\nu=\pm1}\frac{1}{k_{\nu}L\sinh(k_{\nu}L)}\times \nonumber \\
\left[\frac{1}{\sqrt{(\omega_n +iH_{\nu})^2+|\Delta|^2}}-\frac{1}{\sqrt{(\omega_n -iH_{-\nu})^2+|\Delta|^2}}\right]^2
\end{eqnarray}
and  $k_{\nu}=\sqrt{2(\omega_n+i\nu \delta\mu)/D_\text{f}}$, with
$H_{\nu}=\epsilon_{xc}+\nu \delta\mu$ and
$2\delta\mu=\mu_{\uparrow}-\mu_{\downarrow}$.
\begin{figure}[tp]
\includegraphics[width=8cm]{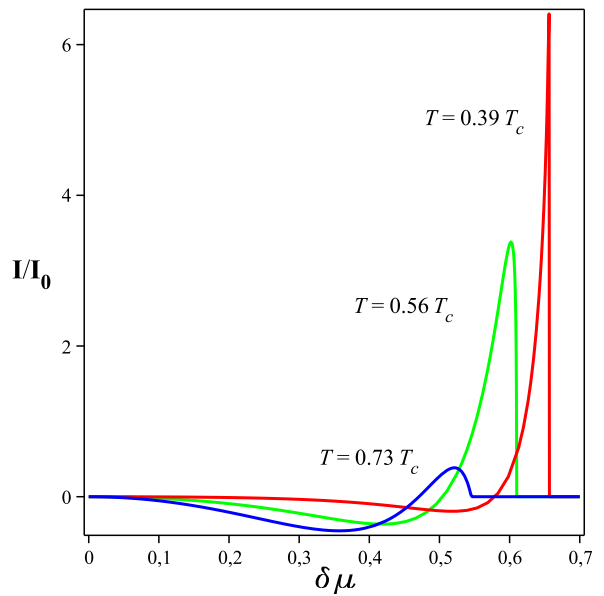}
\caption{(Color online)  The Josephson current as a function of the
spin potential $\delta \mu$ measured in units of the unperturbed
superconducting gap $\Delta_0$, at $L=\zeta$, where
$\zeta=\sqrt{D_\text{f}/2|\Delta_0|}$ and $I_0=(2\zeta\pi k_B
T/er^2_{s\text{f}} \sigma_f)\times$10$^{-2}$}
\end{figure}

\section{Discussion}
As follows from Eq. (\ref{js2}), the Josephson current depends on
the directions of the nonequilibrium spin polarizations in the
superconducting leads. The current reaches its maximum when the spin
accumulations in the leads are perpendicular to the magnetization in
the ferromagnet, $\theta_R=\theta_L=\pi/2$. It reverses its sign
when the spin polarization in one of the leads flips its  direction.
Therefore, in the setup shown in Fig. 1, the junction can be
switched into the $\pi$-state by simply reversing the electric
current through one of the FN contacts. It should be noted
that the dissipative current given by Eq.(\ref{jd}) vanishes when
the relative angles are such that the supercurrent reaches its maximum. Hence,
the dissipative transport can be turned off, a feature that can be important
for practical purposes. Eqs.\ (\ref{js2}) and (\ref{K}) also imply
that the long-range proximity effect, described via $j_s$, vanishes
when the exchange interaction $\epsilon_{xc}=0$. The dependence of
$j_s$ on the spin-potential $\delta \mu$ is shown in Fig. 2. A
finite spin-potential causes variations of the order parameter
$\Delta$ and spin density entering in Eq. (\ref{js2}) which have
been found from a pair of self-consistent equations. In our
calculation of $\Delta$ and $S$, we neglected the exchange field
$\epsilon_{xc}$, assuming that $\epsilon_{xc} \ll \delta\mu$. This
is a realistic assumption, taking into account that  $S < \delta\mu
N_F$ in Eq. (\ref{exc}) and $|\mathrm{G}|$ is considerably less than
1 in some superconducting metals (e.g. Al). In this limit, the
dependence of $\Delta$ on $\delta\mu$ is formally the same as in a
thermally equilibrium superconductor subject to a Zeeman splitting
equal to $\delta\mu$. Such a scenario is well studied in the
literature (see e.g. \cite{Izyumov}). In Fig. 2, we see that the
critical current changes sign at some values of $\delta\mu$. This is
caused by injection of nonequilibrium spins into the ferromagnetic
layer. As a result, the distribution function in Eq. \ref{js} is
different from the equilibrium distribution. There is some
similarity of this effect with a current reversal observed in
Josephson transistors \cite{Bulaevskii}. At the lower temperature,
the supercurrent versus spin-potential is more peaked  in the range
of higher $\delta\mu$, since the spin density increases sharply
together with $\epsilon_{xc}$ in this range. At even higher
$\delta\mu$, the superconductivity is destroyed by spin injection.
That causes a sudden drop of the current. We believe that
this narrow range can be easily observed experimentally in the set
up shown in Fig. 1, because $\delta\mu$ can be fine tuned by varying
the current through the normal leads.

Fig. 2 is calculated at $L=\sqrt{D_\text{f}/2|\Delta_0|}$, that is
the characteristic length of the $\pm 1$-triplet proximity effect in
the range of temperatures considered. This length is obviously much
larger than the s-wave Cooper pair penetration depth
$\sqrt{D_\text{f}/B_{\text{ex}}}$ and therefore clearly demonstrates
how the range of the proximity effect becomes much longer by spin
injection into the superconducting leads.

In conclusion, spin injection into s-wave superconductors can
dramatically increase the stationary Josephson current in SFS
system. This enhancement is provided by $\pm 1$ triplet components
of the electron  pairing function. They are generated in
superconducting leads by exchange fields that are noncollinear with
the ferromagnet magnetization. These fields, in turn, are induced by
an injected spin polarization. Besides a strong effect on the
Josephson current, spin injection also gives rise to a dissipative
current that at zero bias potential is induced due to spin
dependence of the ferromagnet conductivity. Both Josephson and
dissipative currents can be manipulated by varying the injected spin
directions in the leads enabling control of $\pi$-junctions.

Anatoly Mal'shukov gratefully acknowledges  the hospitality of NTNU.

\end{document}